\documentclass[a4paper]{jpconf}
\usepackage{graphicx}
\newcommand{\beq}{\begin{equation}}
\newcommand{\eeq}{\end{equation}}
\newcommand{\aaa}{\hspace{0.3cm}}
\newcommand{\pbar}{\mbox{$\bar{\rm p}$}}
\begin{document}
\begin{flushright}
{\bf LAPTH--Conf--1132/05}
\end{flushright}
\title{Production of antimatter in the galaxy}
\author{P Salati}
\address{LAPTH, 9 Chemin de Bellevue, BP110, 74941 Annecy--le--Vieux Cedex, FR}
\ead{salati@lapp.in2p3.fr}

\begin{abstract}
The astronomical dark matter could be made of weakly interacting massive
species whose mutual annihilations should produce antimatter particles
and distortions in the corresponding energy spectra. The propagation of
cosmic rays inside the Milky Way plays a crucial role and is briefly
presented. The uncertainties in its description lead to considerable
variations in the predicted primary fluxes. This point is illustrated with
antiprotons. Finally, the various forthcoming projects are rapidly reviewed
with their potential reach.
\end{abstract}

\section{Cosmic ray propagation throughout the Milky Way}
\label{sec:propagation}

Supersymmetric neutralinos or Kaluza--Klein particles could explain the dark
matter that is indirectly observed around galaxies, inside their clusters and
on cosmological scales. These species are hunted for by several experimental
collaborations. As they annihilate inside the Milky Way halo, they produce
high--energy cosmic rays and in particular antimatter particles such as antiprotons,
antideuterons and positrons. These are rare elements which are already manufactured
inside the galactic disk through the spallations of primary cosmic rays on the
interstellar gas. That conventional process provides the natural background
inside which the DM induced antimatter signature may be burried and whose
spectral distortions could reveal an exotic signal. It is therefore crucial
to derive as precisely as possible the energy spectra of the various antimatter
species irrespective of the production mechanisms and to evaluate how well
they are known. To reach that goal, a precise understanding of cosmic ray
propagation is mandatory. We briefly sketch here how that propagation is
understood and how well it is modeled.

Interstellar nuclei are accelerated by the passage of supernovae induced
shock waves. The sources for primary cosmic ray nuclei are therefore located
inside the disk of the Milky Way. Primaries propagate then inside the galactic
magnetic fields and undergo collisions on its irregularities -- the Alfv\'en
waves. This motion is described in terms of a diffusion process whose
coefficient $K(E) \; = \; K_{0} \, \beta \, {\mathcal R}^{\delta}$ increases
with the rigidity ${\mathcal R}$ of the cosmic ray particle as a power law.
Because the scattering centers move with a velocity
$V_{a} \sim$ 20 to 100 km s$^{-1}$, a second order Fermi mechanism yields
some diffusive reacceleration whose coefficient $K_{EE}$ may be expressed as
\beq
K_{EE} \; = \;
{\displaystyle \frac{2}{9}} \, V_{a}^{2} \,
{\displaystyle \frac{E^{2} \beta^{4}}{K(E)}} \;\; .
\eeq
Ionization, adiabatic and Coulomb energy losses must also be taken into account
through the energy loss rate $b^{\rm loss}(E)$. Finally, galactic convection
could wipe cosmic rays away from the disk with a velocity
$V_{C} \sim$ 5 to 15 km s$^{-1}$. The number of particles par unit of volume
and space $\psi = dn / dE$ may be derived form the master equation
\beq
V_{C} \, \partial_{z} \Psi \; - \; K \, \Delta \Psi \; - \;
\partial_{E} \left\{ b^{\rm loss}(E) \, \Psi \; + \;
K_{EE}(E) \, \partial_{E} \Psi \right\} \; = \; q \;\; ,
\label{master_equation}
\eeq
that applies for any cosmic ray species -- primary or secondary -- as long as
the various production processes are appropriately described in the rate $q$.
That general scheme may be implemented either
numerically~\cite{Strong:1998pw,Moskalenko:2004fe} or semi--analytically through
the two--zone model discussed in~\cite{Usine1}.

The latter method allows an easier derivation of the theoretical uncertainties
associated to the various parameters at stake -- namely $K_{0}$, $\delta$, $V_{a}$,
$V_{C}$ and the thickness $L$ of the confinements layers that extend above and
beneath the Milky Way disk. The space of these propagation parameters has been
extensively scanned~\cite{Usine1} in order to select the allowed regions where
the predictions on B/C -- a typical CR secondary to primary ratio -- match the
observations. Tens of thousands of propagation models have survived that crucial
test. The diffusion parameters are indeed still largely undetermined contrary to
what is generally believed in our community.
Recently, the same conclusion has been independently reached~\cite{Lionetto:2005jd}
with the help of a fully numerical code~\cite{Strong:1998pw} in which the convective
wind $V_{C}$ increases linearly with vertical heigh $z$. However, the B/C ratio
could not be accounted for when galactic convection and diffusive reacceleration
were both implemented at the same time.

\section{The various antimatter backgrounds and signals}

\subsection{Antiprotons}

\begin{figure}[h!]
{\centerline{\includegraphics[width=24pc]{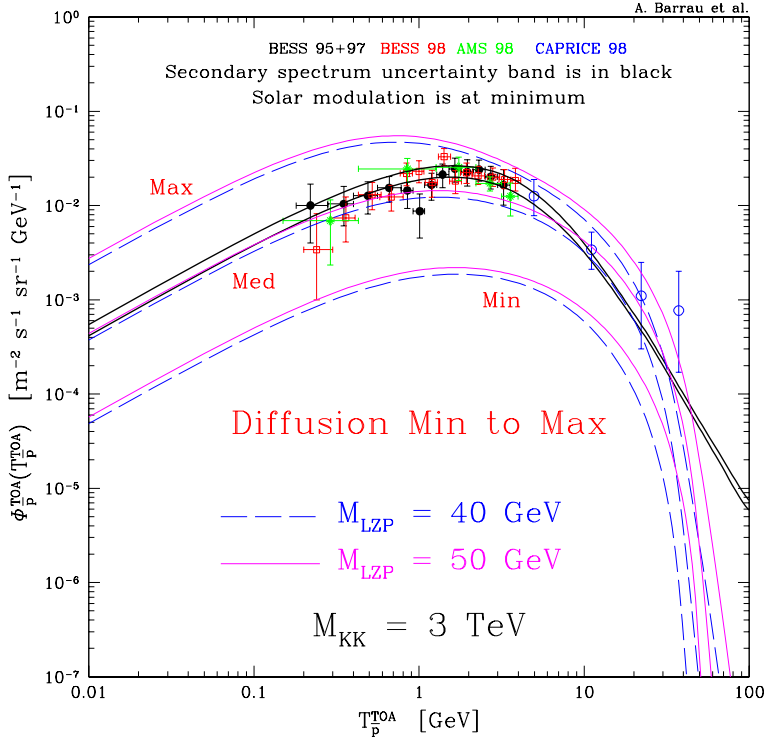}}}
\vskip -0.3cm
\caption{
When the diffusion parameters are varied over the entire domain that is
compatible with the B/C ratio, antiproton primary fluxes span two orders
of magnitude whilst the secondary component lies within a much narrower
band. The case of a resonant LZP~\cite{Agashe2} has been featured
here~\cite{Barrau:2005au} with $M_{\rm LZP} = 40$ -- blue dashed -- and
50 GeV -- solid magenta. A canonical isothermal DM distribution has been
assumed. In the case of minimal diffusion, the LZP signal is well below
the background.}
\label{fig:LZP_diffusion}
\end{figure}

\begin{figure}[h!]
{\centerline{\includegraphics[width=24pc]{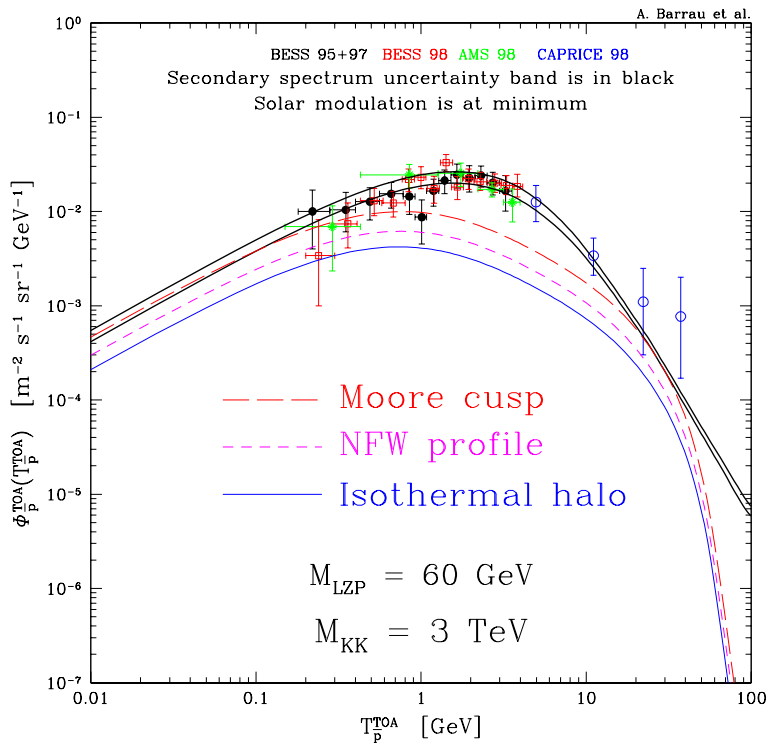}}}
\vskip -0.3cm
\caption{
The effect of the DM halo profile is presented~\cite{Barrau:2005au}
in this figure where the mass of the LZP~\cite{Agashe2} has been set
equal to $M_{\rm LZP} = 60$ GeV with a Kaluza--Klein scale $M_{\rm KK}$
of 3 TeV. The more divergent and concentrated the LZP distribution at
the center of the Milky Way, the larger the antiproton yield. That
effect is particularly acute in this plot where maximum diffusion
parameters have been assumed.}
\label{fig:LZP_halo}
\end{figure}

Secondary antiprotons are produced through the spallations of cosmic ray protons
on the interstellar material
\beq
p \, ({\rm CR}) \; + \; H \, ({\rm ISM}) \;\; \to \;\; \pbar \; + \; X \;\; .
\eeq
The corresponding source rate may be expressed as
\beq
q_{\bar{\rm p}}^{\rm sec}(r , E_{\bar{\rm p}}) \; = \;
{\displaystyle \int_{E^{0}_{\rm p}}^{+ \infty}} \,
{\displaystyle \frac{d \sigma_{\rm p H \to \bar{p}}}{dE_{\bar{\rm p}}}}
\left\{ E_{\rm p} \to E_{\bar{\rm p}} \right\} \; n_{\rm H} \;
v_{\rm p} \; \psi_{\rm p} (r,E_{\rm p}) \;
d E_{\rm p} \;\; ,
\eeq
and can easily be modified in order to incorporate heavier nuclei. The
top--of--atmosphere antiproton flux has been derived~\cite{Usine2} for
each of the numerous propagation models that are compatible with the B/C
measurements~\cite{Usine1}. In spite of the above--mentioned large
uncertainties on the propagation parameters, the predictions all lie
within the two solid black lines of Figures~\ref{fig:LZP_diffusion}
and \ref{fig:LZP_halo} and are in good agreement with the observations
-- a conclusion that has also been reached in~\cite{Lionetto:2005jd}
where a slightly larger uncertainty band is nevertheless obtained.
The estimate of the energy spectrum of secondary antiprotons is therefore
quite robust and does not depend much on the numerous and poorly known
parameters that account for cosmic ray galactic propagation. Actually
since boron and secondary antiprotons both originate from the same
production site -- Milky Way disk -- it is not surprising if the
former species constrains the latter.

On the contrary, primary antiprotons are produced by the mutual
annihilations of DM particles
\beq
\chi \, + \, \chi \;\; \to \;\; q \, + \, \bar{q} \; , \;
W^{+} \, + \, W^{-} \; , \; Z^{0} \, + \, Z^{0} \; , \; H \, + \, H
\;\; \to \;\; \pbar \; + \; X \;\; ,
\eeq
that take place all over the Milky Way DM halo, in a region that
encompasses the above--mentioned confinement layers whose thickness
$L$ is still unknown. As a result, the production of primary antiprotons
whose rate may be expressed as
\beq
q_{\bar{\rm p}}^{\rm DM}(r , E_{\bar{\rm p}}) \; = \;
{\displaystyle \frac{1}{2}} \,
\langle \sigma_{\rm ann} v \rangle \,
{\displaystyle \frac{dN_{\pbar}}{dE_{\pbar}}} \,
\left\{ {\displaystyle \frac{\rho_{\chi}(r)}{m_{\chi}}} \right\}^{2}
\;\; ,
\eeq
suffers from large uncertainties. Predictions span a range of $\sim$
two orders of magnitude at low energy~\cite{pbar_susy,Barrau:2005au}.
This crucial point is illustrated in Figure~\ref{fig:LZP_diffusion}
where the case of a resonant LZP is featured for the maximal, median
and minimal configurations of table~\ref{table:prop}.
%
\begin{table}[h!]
\caption{ Astrophysical parameters of the cosmic ray galactic
propagation models giving the maximal, median and minimal primary
antiproton fluxes compatible wih B/C analysis~\cite{pbar_susy}.
\label{table:prop}}
\begin{center}
{\begin{tabular}{@{}ccccc@{}}
\br
{\rm case} &  $\delta$  & $K_0$                 & $L$   & $V_c$    \\
           &            & [${\rm kpc^{2}/Myr}$] & [kpc] & [km/sec] \\
\mr
{\rm max} &  0.46  & 0.0765 & 15 & 5    \\
{\rm med} &  0.70  & 0.0112 & 4  & 12   \\
{\rm min} &  0.85  & 0.0016 & 1  & 13.5 \\
\br
\end{tabular}}
\end{center}
\end{table}
%
Another source of uncertainty is related to the DM distribution itself
$\rho_{\chi}(r)$ whose radial profile is generically given by
\beq
\rho_{\chi}(r) \; = \; \rho_{\, {\rm DM} \, \odot} \,
\left\{
{\displaystyle \frac{r_{\odot}}{r}} \right\}^{\gamma} \,
\left\{
{\displaystyle \frac{1 \, + \, \left( r_{\odot} / a \right)^{\alpha}}
{1 \, + \, \left( r / a \right)^{\alpha}}}
\right\}^{\left( \beta - \gamma \right) / \alpha} \;\; .
\label{eq:profile}
\eeq
The distance of the Solar System from the galactic center is
$r_{\odot} = 8$ kpc. The local -- Solar System -- DM density has been
set equal to $\rho_{\, {\rm DM} \, \odot} = 0.3$ GeV cm$^{-3}$. In the case
of the pseudo-isothermal profile, the typical length scale $a$ is the radius
of the central core. The profile indices $\alpha$, $\beta$ and $\gamma$ of
the dark matter distributions which have been considered in
Figure~\ref{fig:LZP_halo} are indicated in table~\ref{tab:indices}.
%
\begin{table}[h!]
\caption{ Dark matter distribution profiles in the Milky Way.
\label{tab:indices}}
\begin{center}
{\begin{tabular}{@{}lcccc@{}}
\br
Halo model & $\alpha$ & $\beta$ & $\gamma$ & $a$ [kpc] \\
\mr
Cored isothermal~\cite{bahcall}
& {\aaa} 2 {\aaa} & {\aaa} 2 {\aaa} & {\aaa} 0 {\aaa} & {\aaa} 4
{\aaa} \\
Navarro, Frenk \& White~\cite{nfw}
&        1        &        3        &        1        &
25       \\
Moore~\cite{moore}
&        1.5      &        3        &        1.3      &
30       \\
\br
\end{tabular}}
\end{center}
\end{table}
%
Notice finally that the halo could be clumpy so that the DM species
annihilations could be enhanced by a boost factor that is still very
uncertain.

Constraining the MSSM parameter space or deriving informations on Kaluza--Klein
particles from antiproton measurements may turn out to be tricky. However in
the case of light neutralinos whose relic density accounts for the WMAP
measurement, stringent limits may be extracted. In particular should $L$ exceed
3 kpc, a neutralino lighter than 40 GeV would be basically excluded
-- see~\cite{Bottino:2005xy} for details.

\subsection{Antideuterons}

Secondary antideuterons are manufactured by CR nuclei spallations in which
an antiproton--antineutron pair is produced and then merges. This fusion
is a very rare event because two antinucleons must be produced at the same
time and the difference between their momenta must not exceed $P_{\rm coal}$.
The typical value that has been derived in~\cite{Donato:1999gy} and used
in~\cite{Barrau:2002mc} for that coalescence momentum is $P_{\rm coal} = 58$ MeV.
Because antideuterons are easily destroyed when they interact -- their binding
energy is only 2.2 MeV -- they should not a priori undergo inelastic and
non--annihilating scattering. In the case of antiprotons, that reaction is very
efficient and flattens the spectrum at low energy -- hence the difficulty to
disentangle a primary signal from the secondary background. But in the case
of antideuterons, that process has been assumed until recently to be negligible.
If so, the secondary antideuteron spectrum exhibits a deficiency at low
energy as advocated in~\cite{Donato:1999gy} and primary antideuterons should
be hunted for below a kinetic energy of $\sim 3$ GeV/n.

This point has been extensively reanalyzed recently in~\cite{Duperray:2005si}.
If actually the $\Delta$--resonance excitation
\beq
\bar{\rm D} \; + \; {\rm p} \;\; \to \;\; \bar{\rm D} \; + \; \Delta
\eeq
is forbidden on the basis of isospin conservation, the production of pions
is nevertheless possible through
\beq
\bar{\rm D} \; + \; {\rm p} \;\; \to \;\;
\bar{\rm D} \; + \; {\rm p} \; + \; n \, \pi \;\;.
\eeq
The cross section for inelastic and non--annihilating $\bar{\rm D}$ interactions
is therefore small but non--vanishing. This may allow a certain flattening of
the spectrum in the low energy region and could weaken the relevance of
antideuterons as a clear signature for DM species.
An additional source of spectral flattening investigated in~\cite{Duperray:2005si}
is the production of antideuterons through the collisions of already propagating
antiprotons on interstellar H and He. The threshold for manufacturing a single
antineutron is just $7 \, m_{\rm p}$ -- instead of $17 \, m_{\rm p}$ for a pair --
and low energy antideuterons should be more abundant. The final conclusion reached
in~\cite{Duperray:2005si} -- with a more correct value of $P_{\rm coal} = 79$ MeV
-- strengthens the case for antideuterons. Below $\sim 1 - 2$ GeV/n, the flux of
TOA secondary antideuterons does not exceed
$\sim 10^{-7}$ $\bar{\rm D}$ m$^{-2}$ s$^{-1}$ sr$^{-1}$ (GeV/n)$^{-1}$,
a typical value expected for a primary component in the case of annihilating
supersymmetric neutralinos.

As a matter of fact, a detailed analysis of the detection prospects of
supersymmetric dark matter in the mSUGRA framework has been performed
in~\cite{Edsjo:2004pf} for models whose thermal relic density is in the
range favored by current precision cosmological measurements. Direct detection
and searches for antideuterons were found to be the most promising signatures.

\subsection{Positrons}

Energy losses play a crucial role in the propagation of positrons. Because
galactic convection and diffusive reacceleration can be disregarded, the
master equation~(\ref{master_equation}) simplifies into
\beq
- \; \vec{\nabla} \cdot
\left\{ K \left( \vec{x} , E \right) \, \vec{\nabla} \Psi \right\} \; - \;
\partial_{E}
\left\{ b^{\rm loss}(E) \, \Psi \right\} \; = \; q \left( \vec{x} , E \right) \;\; .
\label{master_2}
\eeq
Above a few GeV, positron energy losses are dominated by synchrotron
radiation in the galactic magnetic fields and by inverse Compton
scattering on stellar light and on CMB photons. The energy loss rate
increases with the positron energy $\epsilon = E / {\rm 1 \, GeV}$ as
$
b^{\rm loss}(E) = {\epsilon^{2}}/{\tau_{E}}
$
where the relevant timescale is typically $\tau_{E} = 10^{16}$ s.
The diffusion coefficient $K$ is assumed to be homogeneous and equal to the
galactic averaged value that is used for antiprotons and antideuterons -- see
section~\ref{sec:propagation}. But if those antinuclei can originate from remote
regions \cite{TailletMaurin,MaurinTaillet} and probe a significant portion of
the Milky Way, the high energy positrons that are detected at the Earth cannot
come from far away and the unknown solar neighborhood value should be preferred.
Future high precision measurements on the $^{10}$Be to $^{9}$Be ratio will be
very valuable since the former nucleus is unstable and also originates from the
local bubble~\cite{2002A&A...381..539D}. Equation~(\ref{master_2}) may be solved
either numerically \cite{Strong:1998pw} or with the help of the smart trick
proposed in~\cite{Baltz:1998xv} which consists in translating the positron
energy $E$ into pseudo--time.
%
%
Measurements of the positron flux show evidence for a spectral
excess~\cite{heat_1,heat_2} at energies $\sim$ 10 GeV. That feature is hard
to explain only with secondary positrons whose flux is nevertheless subject
to theoretical uncertainties -- especially below $\sim$ 5 GeV -- as shown
in~\cite{Lionetto:2005jd}. Numerous analysis have been devoted to the explanation
of the putative HEAT excess in terms of annihilating supersymmetric neutralinos
or Kaluza--Klein species. The latter are more promising because they directly
yield electron--positron pairs with a branching ratio $\sim$ 20\%. Large
values for the boost factor or for the annihilation cross section are in
any case necessary.

The PAMELA~\cite{Pearce:2004ep} and
AMS--02~\cite{Demirkoz:2004en,Jacholkowska:2004fe} missions will collect positrons
up to hundreds of GeV and will bring valuable informations on the HEAT excess and
on the possible presence in the Milky Way halo of DM particles. The reach of these
experiments have been studied in~\cite{Hooper:2004bq} where three years of observation
and a boost factor of 5 have been assumed.
In the case of a bino--like neutralino which mostly produces ${\rm b \bar{b}}$
pairs, PAMELA will be sensitive to configurations down to an annihilation cross
section of order a few times $10^{-26}$ cm$^{3}$ s$^{-1}$ whereas AMS--02 will
lower that value by a factor of 4.
If the neutralino is now a higgsino, PAMELA will be sensitive to masses
up to 380 GeV and AMS--02 will improve that limit up to 650 GeV.
If now the galactic DM is made of Kaluza--Klein species, that reach will
become 550 GeV in the case of PAMELA and 1 TeV for AMS--02.
Remember finally that -- as in the case of antiprotons -- these results
are affected by large uncertainties related to the propagation model.

\section{Forthcoming experiments}

The forthcoming satellite missions PAMELA and AMS--02 have just been mentioned.
They are compared in table~\ref{tab:missions} with the BESS--polar
experiment~\cite{Yoshida:2003ec}.
Long duration flights around the poles where the geomagnetic shield is no longer
active allow for large values of the acceptance. Below 10 GeV, the effective
sensitivity of BESS--polar for low energy antiprotons will be actually 3 times
larger than for PAMELA.
In the same spirit, the CREAM collaboration~\cite{Seo:2003eh}
will measure the B/C ratio between 1 and $10^{3}$ TeV. It aims at the determination
of the index $\delta$ that drives the energy dependence of the diffusion coefficient
$K$. Such a measurement is mandatory in order to reduce the large uncertainties
that plague the theoretical predictions on cosmic rays~\cite{Castellina:2005ub}
and in particular those on the DM induced primaries.
The search for antideuterons is also extremely active. A new upper limit of
$1.9 \times 10^{-4}$ m$^{-2}$ s$^{-1}$ sr$^{-1}$ (GeV/n)$^{-1}$ has been
derived~\cite{Fuke:2005it} for the differential TOA flux of cosmic ray
antideuterons between 0.17 and 1.15 GeV/n. A pioneering
concept~\cite{Mori:2001dv,Hailey:2005yx} for detecting low energy antideuterons
is based on their capture by atoms whose electrons are rapidly expelled.
The result is an antideuteron that orbits alone around the nucleus. As it
cascades towards the ground state, several well defined X--ray photons are
emitted and yield a unique signature as explained by C.~Hailey in this
conference.

%
\begin{table}[h!]
\caption{ Comparison between PAMELA, BESS--polar and AMS--02 as
presented in~\cite{Yoshida:2003ec}. \label{tab:missions}}
\begin{center}
{\begin{tabular}{@{}llll@{}}
\br
Project & PAMELA & BESS--polar & AMS--02 \\
\mr
Flight vehicle & Satellite & long--duration balloon & ISS \\
Flight duration & 3 years & 10--20 days & 3--5 years \\
Altitude & 300--600 km & 37 km (5 g cm$^{-2}$) & 320--390 km \\
Orbit & 70.4$^{\circ}$ & $\geq$ 70$^{\circ}$S Lat. & 51.7$^{\circ}$ \\
Acceptance & 0.0021 m$^{2}$ sr & 0.3 m$^{2}$ sr & 0.3 m$^{2}$ sr \\
MDR (GV) & 740 & 150 & $\sim$ 1000 \\
Particle identification & TOF/TRD/CAL & TOF/ACC & TOF/TRD/RICH/CAL \\
Number of helium & $4 \times 10^{7}$ & $(1-2) \times 10^{7}$ & $2 \times 10^{9}$ \\
Launch & December 2005 & December 2004 & 2007 \\
\br
\end{tabular}}
\end{center}
\end{table}
%


\ack
Pierre Salati would like to thank the organizers for their generous
hospitality as well as for the inspiring and exciting atmosphere
of this very nice meeting.

\section*{References}
%
%
%
\providecommand{\newblock}{}

\end{document}